\documentstyle[preprint,tighten,aps]{revtex}
\newcommand{\be}{\begin{equation}}
\newcommand{\ee}{\end{equation}}
\newcommand{\ba}{\begin{eqnarray}}
\newcommand{\ea}{\end{eqnarray}}
\newcommand{\non}{\nonumber}

\newcommand{\fr}[2]{\frac{#1}{#2}}
\def\vec#1{{\mbox{\boldmath$#1$}}}

\newcommand{\p}{\mbox{$\vec{p}$}}
\newcommand{\Pv}{\mbox{$\vec{P}$}}
\newcommand{\q}{\mbox{$\vec{q}$}}

\newcommand{\r}{\mbox{$\vec{r}$}}
\newcommand{\n}{\mbox{$\vec{n}$}}
\newcommand{\E}{\mbox{$\vec{\cal E}$}}
\newcommand{\e}{\mbox{$\vec{e}$}}
\newcommand{\gra}{\mbox{$\vec{\nabla}$}}
\newcommand{\ep}{\epsilon}

\newcommand{\lb}{\left (}
\newcommand{\rb}{\right )}
\newcommand{\la}{\left\langle}
\newcommand{\ra}{\right\rangle}

\begin{document}

\draft

\preprint{hep-ph/0103241}

\title{Quantum Electrodynamics of the Helium Atom}

\author{ Alexander Yelkhovsky\thanks{
e-mail:  yelkhovsky@inp.nsk.su}}
\address{ Budker Institute of Nuclear Physics,
and Physics Department, Novosibirsk University,\\
Novosibirsk, 630090, Russia} \maketitle

\begin{abstract}
Using singlet $S$ states of the helium atom as an
example, I describe precise calculation of energy
levels in few-electron atoms. In particular, a
complete set of effective operators is derived which
generates ${\cal O}(m\alpha^6)$ relativistic and
radiative corrections to the Schr\"odinger energy.
Average values of these operators can be calculated
using a variational Schr\"odinger wave function.
\end{abstract}


\section{Introduction}

Singlet states of the helium atom, especially its
ground state, are best suited for precision studies of
the electron-electron interaction at low energies.
Recent measurements of $1^1 S - 2^1 P$ \cite{eik} and
$1^1 S - 2^1 S$ \cite{berg} intervals in helium
reached the precision of about 10 {\it ppb}. Helium
ground state ionization potential (the difference
between ground state energies of the singly charged
ion and of the atom) extracted from those measurements
constitutes
\be
\nu_{\rm exp}^{1S-2P}(1^1 S) =
5\;945\;204\;238\,(45)\;{\rm MHz}
\label{eikema}
\ee
and
\be
\nu_{\rm exp}^{1S-2S}(1^1 S) =
5\;945\;204\;356\,(48)\;{\rm MHz},
\label{bergeson}
\ee
respectively.

Theoretically, the ionization potential can be
calculated as a power series in the fine structure
constant $\alpha$. Leading ${\cal O}(m\alpha^2)$
contribution to the ground state energy is the lowest
eigenvalue $E$ of the nonrelativistic Hamiltonian
\be
H = \sum_a \lb \fr{p_a^2}{2m_a}
  + \alpha \sum_{b>a} \fr{z_a z_b}{r_{ab}} \rb
\ee
entering into the Schr\"odinger equation $H \psi = E
\psi$. Here and below I use the following notations:
$\r_a$ and $\p_a$ are the position and momentum
operators for the particle $a$ with mass $m_a$ and
electric charge $z_a$ (in units of the proton charge).
The relative position of two particles is $\r_{ab} =
\r_a - \r_b$; for any vector $\vec{v}$, $v$ denotes
$|\vec{v}|$. The helium atom consists of two electrons
with masses $m_1=m_2=m$ and charges $z_1=z_2=-1$, and
the nucleus with mass $m_3=M$ and charge $z_3=2$. A
general case of $z_3 = Z$ takes into consideration
helium-like ions with $Z\ll 1/\alpha$. In the
center-of-mass frame, total momentum of the atom
vanishes, $\sum_a \p_a = 0$ so that only two of three
position vectors are independent: $\sum_a m_a \r_a =
0$. In singlet states, spins of the electrons sum up
to zero while the orbital part of the wave function is
symmetric with respect to permutation of the electrons
positions, $\psi(\r_1,\r_2) = \psi(\r_2,\r_1)$.

Relativistic and radiative effects shift the
Schr\"odinger value of the energy by corrections of
higher orders in $\alpha$. In particular, the leading
correction,
\ba
\delta^{(2)} E = &&\la - \sum_a \fr{ p_a^4 }{ 8m_a^3 }
             + \pi Z\alpha \fr{\delta(\r_{13})
             + \delta(\r_{23})}{ 2m^2 }
             + \fr{ \pi\alpha \delta(\r_{12})}{ m^2 }
             \right. \non \\
             && \left.
             - \fr{\alpha}{2} \sum_{a;b>a} \fr{z_a z_b}{m_a m_b}
             \left[ \p_a \fr{1}{r_{ab}} \p_b +
             (\p_a\r_{ab})\fr{1}{r_{ab}^3}(\r_{ab}\p_b)
             \right] \ra,
\label{breitcorr}
\ea
arises as the average value of the Breit perturbation
(see, e.g., \cite{BSbook}) over the nonrelativistic
wave function and is of the order $\alpha^2$ relative
to the Schr\"odinger energy. It is taken into account
in (\ref{breitcorr}) that the total spin of electrons
and the spin of the nucleus are both equal to zero.

A particular class of corrections appears due to the
nucleus structure. There, the most important (and
sufficient to be included at the present level of
accuracy) is the effect of the nucleus charge radius
$R_{\rm N}$,
\be
\delta_{\rm chr} E = \fr{2\pi Z \alpha}{3} R_{\rm N}^2
                 \sum_a \la \delta(\r_{a{\rm N}}) \ra.
\label{chrad}
\ee
Here $\r_{a{\rm N}}$ denotes the position of the
$a^{\rm th}$ electron with respect to the nucleus.

The most recent theoretical result for the helium
ionization potential,
\be
\nu_{\rm th}(1^1 S) = 5\;945\;204\;226\,(91)\;{\rm
MHz},
\label{DM}
\ee
obtained in \cite{dm} includes along with
(\ref{breitcorr}) and (\ref{chrad}) the $m\alpha^5$
order, the leading part of the $m\alpha^6$ order, and
some estimates of higher order contributions. Its
uncertainty is twice as large as that of the
experimental data (\ref{eikema}) and (\ref{bergeson}).
The main source of the uncertainty in (\ref{DM}) is
the yet uncalculated part of the ${\cal O}(m\alpha^6)$
correction.

The present work is the first of two devoted to the
calculation of the helium ionization potential with
${\cal O}(m\alpha^6)$ accuracy. It contains the
analytic part of the calculation and its main result
is a set of effective operators which produce all
${\cal O}(m\alpha^6)$ corrections to singlet $S$
levels of helium atom and low-$Z$ helium-like ions. To
make the presentation self-contained, I also briefly
outline how the lower order correction can be obtained
by the same method. The second paper \cite{ky}
contains numerical results for the average values of
the derived effective operators as well as all other
known contributions to the helium ionization
potential.

The rest of this paper is organized as follows.
Section \ref{fw} describes general features of the
approach. Order $m\alpha^5$ effective operators are
derived in Sec. \ref{ma5}. Sections \ref{hard} and
\ref{soft} are devoted to the ${\cal O} (m\alpha^6)$
effective operators appearing from hard and soft
scales, respectively. The final result of this paper
is presented in the Conclusion.

\section{Framework of the calculation}
\label{fw}

Since the early days of quantum electrodynamics (QED)
the nonrelativistic expansion of an atom's ground
state energy is known to break down at the $m\alpha^5$
order \cite{bethe}. In contrast to the ${\cal
O}(m\alpha^4)$ effective operators whose average
values (\ref{breitcorr}) are completely determined by
the soft $(p \sim 1/r \sim m\alpha)$ scale, the
operators of the next order in momentum-to-mass ratio
are too singular to ensure finiteness of their average
values over the ground state described by the wave
function $\psi$\footnote{See Section \ref{ma5} for
details.}. It means that those operators become
sensitive also to the hard $(p \sim m)$ scale which is
beyond the scope of the nonrelativistic expansion.
Another novel feature of the ${\cal O}(m\alpha^5)$
contribution to the energy is that the very picture of
interaction between particles through a potential
fails: virtual transitions from the atom's ground
state to excited states and a photon become relevant.
Thus, one more scale comes into play -- this
intermediate ultrasoft photon has an order $m\alpha^2$
energy. The most natural way to calculate such a
multi-scale shift of the energy is to divide it into
several pieces each originating from its own scale and
then use simplifying approximations suitable to that
scale. For example, the nonrelativistic expansion is
applicable at the soft scale. Alternatively, one can
neglect bound-state effects at the hard scale. If all
relevant contributions are included, their sum is
independent of the details of the division. Applied to
bound state problems in QED, this idea was first
formulated in Ref. \cite{cl} as the nonrelativistic
quantum electrodynamics (NRQED).

Traditionally, in atomic calculations involving
several scales some auxiliary parameter is introduced
to separate a contribution of the given scale from the
others (see, e.g., Ref. \cite{pach} where such a
scheme is applied to the helium problem). For example,
to divide soft and hard scale contributions one
introduces $\lambda$ satisfying $m\alpha \ll \lambda
\ll m$ and gets a final $\lambda$-independent energy
shift as a result of cancellation between two
$\lambda$-dependent contributions. The soft scale one
includes $\lambda$ as the ultraviolet cutoff which
makes average values of singular operators finite.
Simultaneously, $\lambda$ cuts off otherwise infrared
divergent on-shell scattering amplitudes which
represent a hard scale contribution. The higher is the
order of $\alpha$, the more severe are singularities
of both contributions and the larger is a number of
$\lambda$-dependent terms to be canceled in order to
get a final result. The problem seems even less
tractable when the wave function is known only
numerically.

Precise calculations of the positronium spectrum
\cite{ps98,ps99,cmy99,cmy99A} have shown that
contributions from various scales can be separated
much more effectively by shifting the number of
spatial dimensions $d$ from three, $d \to 3-2\ep$. For
consistency, the number of space-time dimensions in
hard scale calculations should be shifted from four to
$4 - 2\ep$. This shift implies essentially that all
objects defined originally for $d=3$ are analytically
continued to the complex plane of $d$. The main
advantage of the dimensional regularization over the
traditional scheme is that due to the analytic
continuation all power divergences automatically drop
out of calculations and one only has to keep track of
logarithmic divergences which show up as inverse
powers of $\ep$.

Recall that there are two kinds of effective operators
in NRQED. Operators coming from the hard scale are
contact, i.e., they are proportional to
delta-functions of distances between particles.
Infrared divergences typical in hard scale
contributions manifest themselves as inverse powers of
$\ep$ in coefficients of those delta-functions. On the
other hand, the soft scale effective operators have
finite coefficients at $d \to 3$. Ultraviolet
divergences inherent to soft scale contributions show
up as inverse powers of $\ep$ only when one evaluates
average values of those operators over a solution of
the Schr\"odinger equation in $d = 3 - 2\ep$
dimensions. The crucial observation made in Refs.
\cite{cmy99,cmy99A} for the ${\cal O}(m\alpha^6)$
corrections to positronium levels is that even without
knowing an explicit form of this solution but using
only the Schr\"odinger equation itself, one can
extract all the divergent pieces in the form of $\la
\delta(\r) \ra/\ep$, where $r$ is a distance between
the electron and the positron while the average value
is calculated over the $d$-dimensional wave function.
Since after such an extraction the divergences
contained in both hard and soft scale contributions
have exactly the same form it is easy to check that
they cancel each other so that a finite remainder can
be safely calculated in three dimensions.

I employ the same idea for the helium atom, where an
analytic form of the wave function is not available
even in three dimensions. Nevertheless, in perfect
analogy to the positronium case, the use of the
Schr\"odinger equation alone makes it possible to
extract the divergent pieces of all soft scale
contributions on the {\em operator} level. Performing
such an extraction I manage to demonstrate
straightforwardly that the divergences coming from
both scales cancel each other {\em before} any
numerical calculation. As the result, the total ${\cal
O}(m\alpha^6)$ correction to a singlet $S$ level is
represented as a sum of apparently finite average
values of the regularization-independent operators.
These average values can be calculated using a wave
function of the helium atom, built as a numerical
solution of the Schr\"odinger equation \cite{ky}.

It is worth mentioning that the idea of the approach
has a simple physical reason. In fact, soft scale
divergences in bound state energy are of the
ultraviolet origin. Hence they should be proportional
to a value of the corresponding wave function at
zeroth separation between interacting particles. In
terms of the effective theory it means that by virtue
of the Schr\"odinger equation one should be able to
rewrite the singular soft scale contributions in such
a way that corresponding divergences are shifted to
the Wilson coefficients of the contact operators.
After the perturbation theory is reformulated in such
a manner, and if the underlying theory is
renormalizable, all divergences that appear in any
given Wilson coefficient have to cancel each other.

Validity of the results obtained below for the helium
can be checked in two limiting cases. The first,
$\alpha \to 0$ at finite $Z\alpha$ describes
helium-like ion with the electron-electron interaction
switched off. The second, $Z \to 0$ at $z_2 \to 1$
describes parapositronium. Since in both cases
three-dimensional wave functions of all $S$ states are
available in an analytic form, the average values of
effective operators can be calculated explicitly
(modulo $\la \delta(\r) \ra/\ep$ terms). Comparison
with the known results shows complete agreement for
all contributions.

In order to make the formulae more transparent, I
write the nonsingular soft scale operators with
coefficients taken at $d=3$.

\section{Order $\lowercase{m}\alpha^5$ effective operators}
\label{ma5}
This Section illustrates the general scheme by the
calculation of effective operators in the first
non-trivial order. As previously mentioned, at ${\cal
O}(m\alpha^5)$ the relevant contributions to the
energy come from three scales: ultrasoft, soft and
hard. Below we will calculate corresponding effective
operators.

\subsection{Hard scale contribution}

Hard scale effects in the interaction between
nonrelativistic particles $a$ and $b$ give rise to the
contact operators which show up as
$c_{ab}\delta(\r_{ab})$ in the spatial representation
and therefore as $c_{ab}$ in the momentum one. In the
NRQED approach, $c_{ab}$ is extracted through the
matching procedure, namely, equating the $ab \to ab$
scattering amplitude calculated in the full QED to
that in the effective theory approach. Dimensional
regularization is best suited to this procedure:
$c_{ab}$ equals minus QED scattering amplitude for the
particles $a$ and $b$ taken on their mass shells and
at rest. In this manner the hard one-loop vertex
correction to the single Coulomb exchange between the
spin-$1/2$ point particle $a$ and a particle $b$
induces the effective potential\footnote{To simplify
the presentation, I omit the factor $(4\pi)^\ep
\Gamma(1+\ep)$ from the final expressions for all
operators in Sect. \ref{ma5}. This factor does not
contribute to the finite total ${\cal O}(m\alpha^5)$
energy correction.}
\be
V_{{\rm h}C}(\r_{ab}) = \fr{2\alpha^2}{3} \fr{z_a^3
z_b}{m_a^2}
                        \lb \fr{1}{\ep} - 2\ln m_a \rb
                        \delta(\r_{ab}).
\label{hc}
\ee
The electron's anomalous magnetic moment taken in the
leading one-loop approximation gives rise to the
potential between electrons $a$ and $b$
\be
V_{\rm hm}(\r_{ab}) = - \fr{8\alpha^2}{3m^2}\,
                        \vec{s}_a \vec{s}_b\,
                        \delta(\r_{ab}),
\label{hm}
\ee
where $\vec{s}_a$ is the spin operator of the $a^{\rm
th}$ particle. If the nucleus spin is zero, the
corresponding electron-nucleus potential vanishes. The
next effective potential is generated by the hard
one-loop box diagrams. For two spin-$1/2$ particles
such a potential reads \cite{ps98D}
\be
V_{\rm box}(\r_{ab}) = \fr{(\alpha z_a z_b)^2}{ m_a
m_b }
              \lb \fr{1}{\ep} - \ln (m_a m_b) - \fr{1}{3}
              + \fr{m_a + m_b - 2\mu_{ab}
              (1+4\vec{s}_a \vec{s}_b)}{m_a - m_b}
              \ln \fr{m_a}{m_b}  \rb \delta(\r_{ab}),
\label{box}
\ee
where $\mu_{ab} = m_a m_b/(m_a + m_b)$ is the reduced
mass of the pair. With the ${\cal O}(m/M)$ precision,
the corresponding electron-nucleus effective potential
is
\be
V_{\rm box}(\r_{\rm eN}) = \fr{(Z\alpha)^2}{m M}
           \lb \fr{1}{\ep} - 2\ln m - \fr{1}{3} \rb
           \delta(\r_{\rm eN}).
\ee
The last ${\cal O}(m\alpha^5)$ contribution coming
from the hard scale appears due to vacuum
polarization. In ordinary few-electron atoms, an
account of the electron vacuum polarization is
sufficient for the present-day accuracy:
\be
V_{\rm vp}(\r_{ab}) = \fr{4\alpha^2}{15}
               \fr{z_a z_b}{m^2}  \delta(\r_{ab}).
\label{vp}
\ee
The hard scale contribution to the energy equals the
average value of $V_{{\rm h}C}+V_{\rm hm}+V_{\rm
box}+V_{\rm vp}$ summed over all pairs of particles.

\subsection{Ultrasoft scale contribution}

According to the standard rules of the perturbation
theory, virtual transition of the atom into an excited
state induced by the emission and subsequent
absorption of a photon is described by the following
effective operator:
\be
\int \fr{d^d q}{(2\pi)^d} \fr{4\pi\alpha}{2q}
     \sum_a j_{ai} \exp(-i\vec{q}\vec{r}_a)
     \fr{\delta_{ij} - \fr{q_i q_j}{q^2}}{E - H - q}
     \sum_b j_{bj} \exp(i\vec{q}\vec{r}_b),
\label{usinit}
\ee
where $H$ is the $d$-dimensional Schr\"odinger
Hamiltonian, $E$ is its lowest eigenvalue, $\q$ is the
photon momentum, $q = |\q|$. Assuming that $q \sim
m\alpha^2$, we can restrict our attention to the
electric dipole transitions, i.e. replace the
exponents $\exp(\pm\q\r_a)$ by 1 and the current
density operators $\vec{j}_a$ by their orbital
counterparts taken in the leading nonrelativistic
approximation, $\vec{j}_a \to z_a\p_a/m_a$, in
(\ref{usinit}). Integration over directions of $\q$
then gives for the ultrasoft scale effective operator
\be
V_{\rm us} = \fr{\alpha\Omega_d}{(2\pi)^{d-1}}
\fr{d-1}{d}\,
             \vec{J}
             \int_0^\infty \fr{dq\, q^{d-2}}{E-H-q}\;
             \vec{J},
\label{usd}
\ee
where the operator $\vec{J}$ is defined as $\sum_a z_a
\p_a /m_a$ and $\Omega_d = 2\pi^{d/2}/\Gamma(d/2)$ is
the $d$-dimensional solid angle. Analytic continuation
of the integral over $q$ from the stripe $1 < {\rm
Re}(d) < 2$ reads
\be
\int_0^\infty \fr{dq\, q^{d-2}}{E-H-q} =
              \fr{\pi (H-E)^{d-2}}{\sin (\pi d\,)}.
\ee
Expanding now the right-hand side of (\ref{usd}) in
$\ep = (3-d)/2$, we get with the ${\cal O}(\ep^0)$
accuracy
\be
V_{\rm us} \to \fr{2\alpha}{3\pi} \vec{J}
           (H-E) \lb \fr{1}{2\ep} + \fr{5}{6} - \ln 2(H-E) \rb
           \vec{J}.
\label{usfin}
\ee
As previously mentioned, the $1/\ep$ term is due to
the divergence of $V_{\rm us}$ in three dimensions.
The ${\cal O}(m\alpha^5)$ ultrasoft scale contribution
to the energy is the average value of the operator
(\ref{usfin}) over the $d$-dimensional wave function:
\be
\delta^{(3)}_{\rm us} E = \fr{2\alpha}{3\pi} \la
              \lb \fr{1}{2\ep} + \fr{5}{6} - \ln(m\alpha^2) \rb
              \fr{[\vec{J},[H,\vec{J}]]}{2} -
              \vec{J} (H-E) \ln \fr{H-E}{{\rm Ry}} \vec{J} \ra.
\label{eusinit}
\ee
Here I used the Schr\"odinger equation and also the
standard notation for the Rydberg constant ${\rm
Ry}=m\alpha^2/2$. Since the Poisson equation
$[\p_a,[\p_a, C_{ab}]] = 4\pi\alpha z_a z_b
\delta(\r_a - \r_b)$ for the Coulomb potential
$C_{ab}$ between two particles holds in any
dimensions, we have:
\be
[\vec{J},[H,\vec{J}]] = -4\pi\alpha \sum_a z_a
  \sum_{b > a} z_b \lb \fr{z_a}{m_a} - \fr{z_b}{m_b} \rb^2
  \delta(\r_{ab}).
\ee
The ultrasoft correction (\ref{eusinit}) in helium
then reads
\be
\delta^{(3)}_{\rm us} E = \fr{4Z\alpha^2}{3}
                    \lb \fr{1}{m} + \fr{Z}{M} \rb^2
              \lb \fr{1}{2\ep} + \fr{5}{6} - \ln(m\alpha^2)
              - \ln \fr{k_0}{{\rm Ry}} \rb
              \la \delta(\r_{23}) + \delta(\r_{31}) \ra,
\label{eus}
\ee
where the helium Bethe logarithm \cite{ks} is defined
as
\be
\ln \fr{k_0}{{\rm Ry}} = \fr{\la \p_3 (H-E)
            \ln \fr{H-E}{{\rm Ry}} \p_3 \ra}{
            \la \p_3 (H-E) \p_3 \ra }
\ee
and can be safely calculated in three dimensions.

\subsection{Soft scale contribution}

At the soft scale, momenta of all particles
(electrons, nucleus and virtual photons) are of the
order $m\alpha$. Therefore the ${\cal O}(m\alpha^5)$
soft scale effective operators are generated by
transverse photon exchange(s) since only such
scattering amplitudes can contain odd powers of photon
momentum.

\subsubsection{Retardation}

Let us again start with the single transverse photon
exchange described by the effective operator
(\ref{usinit}) but now assuming that $q \sim m\alpha$.
Since $H - E \ll m\alpha$, we can expand the integrand
in $(H-E)/q$. Zeroth order term of this expansion
describes the magnetic interaction in instantaneous
approximation and is included (modulo relativistic
corrections) into the Breit perturbation (see
(\ref{breitcorr})). The first order retardation effect
is represented by the operator
\be
V_{\rm ret} = 4\pi\alpha
     \int \fr{d^d q}{(2\pi)^d}
     \fr{\delta_{ij} - \fr{q_i q_j}{q^2}}{2q^3}
     \sum_a j_{ai} \exp(-i\vec{q}\vec{r}_a)
     (H - E)
     \sum_b j_{bj} \exp(i\vec{q}\vec{r}_b).
\label{retinit}
\ee
Using again the Schr\"odinger equation, we get
\be
\delta_{\rm ret} E = \pi\alpha
     \int \fr{d^d q}{(2\pi)^d}
     \fr{\delta_{ij} - \fr{q_i q_j}{q^2}}{q^3}
     \la \left[ \sum_a j_{ai} \exp(-i\vec{q}\vec{r}_a),
     \left[H,\sum_b j_{bj} \exp(i\vec{q}\vec{r}_b) \right]
     \right] \ra.
\label{eret}
\ee
The order $m\alpha^5$ correction arises due to the
nonrelativistic current densities,
\be
\vec{j}_a(\p',\p) \to z_a \fr{\p'+\p +
2[(\p'-\p)\vec{s}_a,\vec{s}_a]}{2m}.
\ee
We then see that in (\ref{eret}) only the exchange
between different particles ($a \neq b$) can give a
nonzero contribution. In fact, the integral over $q$
in the `diagonal' terms ($a = b$) is scaleless and
hence vanishes. The expression (\ref{eret})
simplifies,
\be
\delta^{(3)}_{\rm ret} E = -2\pi\alpha
     \sum_a \sum_{b>a} \fr{z_a z_b}{m_a m_b}
     \la U_{ij}(\r_{ab})
     \left[p_{ai},\left[p_{bj}, C_{ab}(\r_{ab})\right]
     \right] \ra,
\label{eret5}
\ee
and reproduces the result obtained in Ref.
\cite{my-logs} for positronium. Here
\be
U_{ij}(\r)  = \int \fr{d^d q}{(2\pi)^d}
              \fr{\delta_{ij} - \fr{q_i q_j}{q^2}}{q^3}
              \exp(i\q \r) = \fr{\Gamma\lb\fr{d-3}{2}\rb
              r^{3-d} }{6 \pi^{(d+1)/2} }
              \lb \delta_{ij} + \fr{d-3}{2} n_i n_j\rb,
\label{U}
\ee
$C_{ab}(\r) = \alpha z_a z_b \Gamma(d/2-1) r^{2-d}/
\pi^{d/2-1}$ is the Coulomb potential in $d$
dimensions and $\vec{n} = \r/r$. Thus we get
\be
\delta^{(3)}_{\rm ret} E = - \fr{4\alpha^2}{3}
     \fr{\Gamma\lb\fr{d+1}{2}\rb \Gamma\lb\fr{d}{2}\rb}{
         \pi^{d-3/2}}
     \sum_a \sum_{b>a} \fr{z_a^2 z_b^2}{m_a m_b}
     \la r_{ab}^{3-2d} \ra.
\label{eret5int}
\ee
Here we cannot take the limit $d \to 3$ since the
average value of $r^{-3}$ diverges logarithmically.
However, we can extract the divergence in the
following way. By definition, we have
\be
\la r_{ab}^{3-2d} \ra = {\rm a}_{ab}^{4\ep}
    \int d \r' \int d \n_{ab}
    \int_0^{\infty} \fr{ d\rho }{ \rho^{1-2\ep} }
    \psi^2({\rm a}_{ab}\rho\n_{ab},\r').
\ee
Here ${\rm a}_{ab} = |z_a z_b \mu_{ab}\alpha|^{-1} $
is the Bohr radius for a given pair of particles,
$\r_{ab} = {\rm a}_{ab}\rho\n_{ab}$ being their
relative position. The remaining independent variables
are denoted by $\r'$. Integrating by parts in the last
integral, we get:
\be
\int_0^{\infty} \fr{ d\rho }{ \rho^{1-2\ep} }
    \psi^2 ({\rm a}_{ab}\rho\vec{n}_{ab},\r')
    = - \fr{1}{2\ep}
    \int_0^{\infty} d \rho\, \rho^{2\ep}
    \fr{\partial}{\partial \rho}
    \psi^2 ({\rm a}_{ab}\rho\vec{n}_{ab},\r').
\label{singint}
\ee
Here I took into account that $\lim_{\rho\to 0}
\rho^{2\ep}$ being written as the integral over
momentum has no scale and hence vanishes. Substituting
$\rho^{2\ep} \to 1 + 2\ep\ln\rho + {\cal O}(\ep^2)$
into the r.h.s. of (\ref{singint}) gives for the
retardation correction (\ref{eret5int}):
\be
\delta^{(3)}_{\rm ret} E \to - \fr{4\alpha^2}{3}
     \sum_a \sum_{b>a} \fr{z_a^2 z_b^2}{m_a m_b} \left\{
     \lb \fr{1}{\ep} - 2\ln \fr{4}{{\rm a}_{ab}} - 1 \rb
     \la \delta(\r_{ab}) \ra
     - \la \fr{ \gamma + \ln\fr{2 r_{ab}}{{\rm a}_{ab}}
     }{ \pi r_{ab}^2 }
     \n_{ab} \overrightarrow{\gra}_{ab} \ra \right\},
\label{eret5fin}
\ee
where $\gra_{ab} = \partial/\partial \r_{ab}$ and
$\gamma = 0.5772\ldots$ is the Euler constant. The
arrow shows what the gradient is acting on. Thus we
have managed to extract the divergences in the form of
average values of the contact operators
$\delta(\r_{ab})$ divided by $\ep$. Since $\psi$ and
its first derivatives are finite\footnote{It follows
from the Schr\"odinger equation in $d$ dimensions.}
for $\r_{ab} \to 0$, the non-contact average values in
(\ref{eret5fin}) are finite in three dimensions.

\subsubsection{Double seagull}

One more soft scale contribution of the order
$m\alpha^5$ appears due to the double transverse
exchange between two particles when both photons are
emitted and reabsorbed in the seagull vertices. The
corresponding effective potential derived in
\cite{my-logs} for the positronium can be easily
generalized to a more complex atom:
\be
V_{\rm ds} = - 2\alpha^2
\fr{\Gamma(1-\ep)^2}{(4\pi)^{1-2\ep}}
             \left[ 1 - \ep \fr{17 - 8\ln 2}{3}
             + {\cal O}(\epsilon^2) \right]
             \sum_a \sum_{b>a} \fr{z_a^2 z_b^2}{ m_a m_b }
             \la r_{ab}^{3-2d} \ra.
\ee
Exploiting the same trick as above to extract the
divergences we get the double seagull contribution to
the energy:
\be
\delta^{(3)}_{\rm ds} E \to - \alpha^2
     \sum_a \sum_{b>a} \fr{z_a^2 z_b^2}{m_a m_b} \left\{
     \lb \fr{1}{\ep} + 2\ln {\rm a}_{ab} - \fr{11-2\ln 2}{3} \rb
     \la \delta(\r_{ab}) \ra
     - \la \fr{ \gamma + \ln\fr{2 r_{ab}}{{\rm a}_{ab}}
     }{ \pi r_{ab}^2 }
     \n_{ab} \overrightarrow{\gra}_{ab} \ra \right\}.
\label{eds}
\ee

\subsection{Total $\lowercase{m}\alpha^5$ correction}

The $\ep^{-1}$ terms cancel out in the sum of all
${\cal O}(m\alpha^5)$ corrections to the energy. Hence
we can take the limit $d \to 3$ in this sum. With the
${\cal O}(m^2/M^2)$ precision, the result for the
helium ground state reads:
\ba
\delta^{(3)} E &=& -\fr{2 Z\alpha^2}{3m^2}
                   \lb 4\ln(Z\alpha)
                   + 2\ln \fr{k_0}{Z^2{\rm Ry}}
                   - \fr{19}{15} 
                   \rb \la \delta(\r_1) + \delta(\r_2) \ra
                   \non \\
               &&  +\fr{2\alpha^2}{3m^2}
                   \lb 7\ln\alpha + \fr{82}{5} \rb
                   \la \delta(\r) \ra
                   + \fr{7\alpha^2}{3\pi m^2}
                   \la \fr{ \gamma + \ln (m\alpha r)
                        }{ r^2 }
                 \n \overrightarrow{\gra} \ra
                 \non \\
               &&  -\fr{2 (Z\alpha)^2}{3m M}\lb \ln(Z\alpha)
                   + 4 \ln \fr{k_0}{Z^2{\rm Ry}}
                   - \fr{31}{3} \rb
                 \la \delta(\r_1) + \delta(\r_2) \ra
                   \non \\
               && + \fr{7 (Z\alpha)^2}{3\pi m M}
                 \la \fr{ \gamma + \ln (Z m\alpha r_1)
                        }{ r_1^2 }
                 \n_1 \overrightarrow{\gra}_1
                 + \fr{ \gamma + \ln (Z m\alpha r_2)
                        }{ r_2^2 }
                 \n_2 \overrightarrow{\gra}_2 \ra.
\label{e5}
\ea
Here and below I use simplified notations: $\r_1 = r_1
\n_1 = \r_{23}$, $\r_2 = r_2 \n_2 = \r_{31}$, $\r = r
\n = \r_{12}$, the gradients are taken over the
corresponding position vectors, $\gra_1 =
\partial/\partial \r_1 $ and so on. In the limit of no
recoil ($m/M \to 0$), the result (\ref{e5}) agrees
with the results of Araki \cite{araki} and Sucher
\cite{sucher} after integrating by parts in their
average value $Q$:
\be
Q = \lim_{\rho \to 0} \la \fr{\Theta(r - \rho)
                           }{ 4\pi r^3 }
                       + ( \gamma + \ln (m\alpha\rho) )
                         \delta(\r) \ra
  = - \fr{1}{2\pi} \la \fr{ \gamma + \ln (m\alpha r)
                        }{ r^2 }
                        \n \overrightarrow{\gra} \ra.
\ee
The first recoil (linear in $m/M$) correction was
previously discussed in Ref. \cite{pach98}.

\section{Order $\lowercase{m}\alpha^6$ Hard Scale
Contributions}
\label{hard}

Similarly to what was done in the previous order, one
has to consider the hard scale part of a two-particle
scattering amplitude but now in two loops. There is no
need to consider {\em three}-particle scattering
amplitudes. In fact, the probability density to find
three particles forming the helium atom at the same
point is of the order $(m\alpha)^6$. On the other
hand, these particles should exchange at least three
photons to form a hard loop. Hence, hard scale
effective operators proportional to
$\delta(\r_{ab})\delta(\r_{bc})$ can produce an ${\cal
O}(m\alpha^9)$ correction only.

The radiative recoil potential appears when we account
for the first radiative corrections to the hard
one-loop box diagrams (see, e.g., \cite{cmy99A}). The
corresponding two-loop diagrams involve only even
powers of the electric charges $z_1$ and $z_2$. Hence
the radiative recoil effective operator coincides with
that for parapositronium \cite{pk,cmy99A}:
\be
\delta_{\rm rad~rec} E = \lb \fr{6\zeta(3)}{\pi^2}
                    - \fr{697}{27\pi^2}
                    - 8\ln 2 + \fr{1099}{72} \rb
                      \fr{\pi\alpha^3}{m^2}
                      \la \delta(\r) \ra.
\label{radrec}
\ee
The corresponding electron-nucleus operator vanishes
in the non-recoil limit $m/M \to 0$ considered from
now on\footnote{Order $m^2\alpha^6/M$ correction is
much less than unknown $m\alpha^7$ corrections.}.
Then, one- \cite{1loopeN} and two-loop
\cite{slope,pauliff,vacpol} pure radiative corrections
to electron--nucleus interaction give rise to the
following energy shifts:
\be
\delta^{\rm eN }_{\rm rad1l} E =
          \lb \fr{427}{96} - 2\ln 2 \rb
          \fr{\pi\alpha(Z\alpha)^2}{m^2}
          \la  \delta(\r_1) + \delta(\r_2) \ra,
\label{rad1}
\ee
\be
\delta^{\rm eN }_{\rm rad2l} E =
          \lb - \fr{9\zeta(3)}{4\pi^2}
                    - \fr{2179}{648\pi^2}
                    + \fr{3\ln 2}{2} - \fr{10}{27} \rb
          \fr{\pi\alpha^2(Z\alpha)}{m^2}
          \la \delta(\r_1) + \delta(\r_2) \ra.
\label{rad2}
\ee
The net effect of two-loop contributions to the slope
of the electron Dirac formfactor \cite{slope}, Pauli
formfactor \cite{pauliff}, and vacuum polarization
\cite{vacpol} reads:
\be
\delta^{\rm ee}_{\rm rad} E =
          \lb \fr{15\zeta(3)}{2\pi^2}
                    + \fr{631}{54\pi^2}
                    - 5\ln 2 + \fr{29}{27} \rb
          \fr{\pi\alpha^3}{m^2} \la \delta(\r) \ra.
\label{radee}
\ee
Finally, to get the pure recoil contribution to the
electron-electron hard scale interaction (three photon
exchange) we have to change the sign of the
corresponding parapositronium result\footnote{Recall
that it is convenient to omit the overall factor
$(4\pi)^{2\ep} \Gamma^2(1+\ep)$ from the final
expressions for all ${\cal O}(m\alpha^6)$ operators.}
\cite{cmy99A}:
\be
\delta_{\rm rec} E =  \lb - \fr{1}{\epsilon}
                    + 4\ln m
                    - \fr{39\zeta(3)}{\pi^2} + \fr{32}{\pi^2}
                    - 6\ln 2 + \fr{7}{3} \rb
                      \fr{\pi\alpha^3}{4m^2}
                      \la \delta(\r) \ra.
\label{rec}
\ee
Among the hard scale contributions only the last one
contains the divergence.

\section{Order $\lowercase{m}\alpha^6$ Soft Scale
Contributions}
\label{soft}

The aim of this Section is to demonstrate that in
analogy to the previous order the sum of all singular
average values reduces to the form
\be
\delta_{\rm soft}^{\rm sing} E = \fr{1}{\epsilon}
                      \fr{\pi\alpha^3}{4m^2}
                      \la \delta(\r) \ra,
\label{softdiv}
\ee
so that the sum of soft and hard scale contributions
is finite in three dimensions.

There are many soft scale effective operators with
singular average values. One can easily determine
whether an average value of a given operator is
singular or regular for $d \to 3$ using only the fact
that the wave function and its first derivatives are
finite when positions of two particles coincide.

\subsection{Irreducible corrections}

\subsubsection{Dispersion correction}

Let us consider details of the singularities
extraction procedure using the dispersion correction
as an example. Nonrelativistic expansion of the
electron's dispersion law, $\omega_p=\sqrt{m^2+p^2}$,
reads:
\be
\omega_p = m + \fr{p^2}{2m} - \fr{p^4}{8m^3}
           + \fr{p^6}{16m^5} + \ldots.
\ee
The last written term induces a correction of the
appropriate order. Using the Schr\"odinger equation we
get
\ba
\delta_{\rm disp} E &=& \la \fr{ p_1^6+p_2^6 }{ 16m^5
} \ra = \fr{1}{2m^2} \la
      \lb \fr{ p_1^2+p_2^2 }{ 2m } \rb^3
    - 3 \fr{ p_1^2 p_2^2 }{ 4m^2 }
        \fr{ p_1^2+p_2^2 }{ 2m } \ra
     \non \\
&=& \fr{1}{2m^2} \la (E-C)(H-C)(E-C)
    - 3 \left\{ \fr{ p_1^2 p_2^2 }{ 8m^2 },E-C\right\} \ra
     \non \\
&=& \fr{1}{2m^2} \la (E-C)^3
    + \left[ C, \fr{[H,C]}{2}  \right]
    - 3 \fr{ p_1^2 p_2^2 }{ 4m^2 }E
    + \fr{3}{8} \left\{
        \fr{ p_1^2 p_2^2 }{ m^2 },
        C \right\} \ra.
\label{dispinit}
\ea
Here the total Coulomb potential $C$ is the sum of the
electron-nucleus and electron-electron parts,
\be
C = C_{\rm N} + c = C_1 + C_2 + c,
\ee
while a pairwise Coulomb potential is defined after
Eq.(\ref{U}). Singular contributions to
(\ref{dispinit}) are induced by the following
operators: i) $C^3$; ii) the double commutator,
\be
\fr{1}{2m^2} \la \left[ C, \fr{[H,C]}{2}  \right] \ra
= \fr{1}{2m^2} \la \left[ C,
       \sum_{a=1}^2 \fr{ \E_a\gra_a }{ 2m }
       + \fr{ \e\gra }{ m }
       \right] \ra
= \la \fr{ (\E_1-\E_2)\e }{ 2m^3 } + \fr{ {\cal
E}_1^2+{\cal E}_2^2 }{ 4 m^3 }
         + \fr{ e^2 }{ 2 m^3 } \ra,
\label{doublecomm}
\ee
where
\be
\E_a = - \left[ \gra_a, C_a \right]; \qquad \e = -
\left[ \gra, c \right]
\ee
are electric forces exerted on electrons; and iii) the
anticommutator
\ba
\fr{3}{16m^2} \la \left\{
        \fr{ p_1^2 p_2^2 }{ m^2 },
        c \right\} \ra
&=& \fr{3}{16m^2} \la  \left\{
          \lb \fr{ p_1^2+p_2^2 }{ 2m }
          \rb^2
          - \lb \fr{ p_1^2-p_2^2 }{ 2m }
          \rb^2,
          c \right\}\ra
          \non \\
&=& \fr{3}{16m^2} \la 2c(E-C)^2
          - \left[ c, [H,C] \right]
          - \left\{ c,
             \fr{ (\Pv \p)^2 }{ m^2 }
             \right\}\ra,
\label{ac}
\ea
where $\Pv = \p_1 + \p_2$ and $\p = (\p_1 - \p_2)/2$.
Double commutator $\left[ c, [H,C] \right]$ can be
transformed similarly to (\ref{doublecomm}), the last
term from (\ref{ac}) is conveniently rewritten as
\be
- \fr{3}{16m^2} \la \left\{ c,
                \fr{ (\Pv \p)^2 }{ m^2 }
                \right\}\ra =
- \fr{3}{16m^4} \la \left[ \Pv \p,
        \left[ \Pv \p, c \right] \right] +
        2 (\Pv \p) c (\Pv \p) \ra.
\ee
Summing up all of the above contributions and using
the virial theorem in three dimensions, $\la C \ra =
2E$, we get for the dispersion correction:
\ba
\delta_{\rm disp} E &=& - \fr{5E^3}{2m^2}
       + \fr{3 E^2 \la c \ra}{8m^2}
       + \fr{3E}{2m^2} \la C^2 - \fr{C c}{2}
                - \fr{ p_1^2 p_2^2 }{ 4m^2 }
                \ra
              \non \\
     &&
+ \la \fr{ 3 p_1^2 C_{\rm N} p_2^2 }{ 8 m^4 }
         -  3 \fr{ \left[ \Pv \p,
        \left[ \Pv \p, c \right] \right] }{ 16m^4 }
- 3  \fr{ (\Pv\p) c (\Pv\p) }{ 8m^4 }
       - \fr{ 9 C_{\rm N}^2 c }{ 8m^2 }
       - \fr{ 3 C_{\rm N} c^2 }{ 4m^2 }
             \right.  \label{disp}
\\
     &&
  \left. + \fr{ 5(\E_1-\E_2)\e }{ 16m^3 }
         - \fr{ C_{\rm N}^3 }{ 2m^2 }
         - \fr{ c^3 }{ 8m^2 }
         + \fr{ {\cal E}_1^2+{\cal E}_2^2 }{ 4m^3 }
         + \fr{ e^2 }{ 8m^3 } \ra.
         \non
\ea
One can easily check that only the operators $C_a^3$,
$c^3$ and ${\cal E}_a^2$, $e^2$ have divergent average
values in three dimensions. The following analysis
shows that in a similar manner singularities of all
soft scale effective operators appear either as the
third power of the Coulomb potentials or as the
electric forces squared.


\subsubsection{Coulomb corrections}

The order $p^4/m^4$ correction to the Coulomb
potential between electron and nucleus reads:
\be
V_C^{\rm eN} = \fr{ 5 }{ 128m^4 } \left[ p_1^2,
              [ p_1^2, C_1 ] \right]
      - \fr{ 3\pi Z\alpha }{ 16m^4 }
        \left\{ p_1^2, \delta(\r_1) \right\}
      + (1 \leftrightarrow 2).
\label{VCeN}
\ee
I drop the spin-orbit term, which vanishes in a state
with the total spin zero. The average value of the
double commutator can be conveniently rewritten as
\ba
\la \left[ p_1^2, [ p_1^2, C_1 ] \right]
    + (1 \leftrightarrow 2) \ra
&=& 2m  \la \left[ H-C, [ p_1^2, C_1 ]
       + (1 \leftrightarrow 2) \right]  \ra
    \non \\
&=& - 4m \la {\cal E}_1^2 + {\cal E}_2^2 +
(\E_1-\E_2)\e \ra.
\ea
For the anticommutator in (\ref{VCeN}), it should be
taken into account that the average value of the
operator $\delta(\r_a) C_a$ is a scaleless integral
over $\r_a$ and hence vanishes in the dimensional
regularization:
\be
\la \left\{ p_1^2, \delta(\r_1) \right\}
    + \left\{ p_2^2, \delta(\r_2) \right\} \ra
= \la \delta(\r_1) \lb 4m (E - C_2 - c) - 2p_2^2
                   \rb + (1 \leftrightarrow 2) \ra.
\ee
The energy shift induced by (\ref{VCeN}) is
\be
\delta_C^{\rm eN} E = \la \fr{ 3\pi Z\alpha
                     }{ 4m^3 }\delta(\r_1)
                   \lb \fr{ p_2^2  }{ 2m } + C_2 + c
                       - E \rb
               - \fr{ 5 \E_1(\E_1+\e) }{ 32m^3 }  \ra
             + (1 \leftrightarrow 2).
\label{CeN}
\ee
Note that $\e$ changes its sign under the permutation
$(1 \leftrightarrow 2)$.

Similar analysis for the correction to the
electron-electron Coulomb interaction,
\be
V_C^{\rm ee} =
      \fr{ 5 }{ 128m^4 } \left[ p_1^2,
              [ p_1^2, c ] \right]
      + \fr{ 7\pi\alpha }{ 32m^4 }
        \left\{ p_1^2, \delta(\r) \right\}
      + (1 \leftrightarrow 2)
      + \fr{ (\p_1\p_2) c (\p_1\p_2)
            - p_{1i}p_{2j}\; c\; p_{1j}p_{2i} }{ 16m^4 },
\label{VCee}
\ee
gives
\ba
\delta_C^{\rm ee} E = \la 3  \fr{ \left[ \Pv \p,
        \left[ \Pv \p, c \right] \right] }{ 32m^4 }
                  + \fr{ \pi\alpha\delta(\r) }{ m^3 }
                    \lb E - C_{\rm N} - \fr{3 P^2 }{ 32m }
                    \rb
                  - \fr{ (\E_1-\E_2)\e }{ 16m^3 }
                  - \fr{ e^2 }{ 8m^3 }\ra.
\label{Cee}
\ea
Virtual transitions of electrons to negative-energy
states induced by the Coulomb exchanges generate the
energy shift
\be
\delta_{C}^- E = \la \fr{ {\cal E}_1^2+{\cal E}_2^2
                         + 2(\E_1-\E_2)\e
                         + 2 e^2
                        }{ 8m^3 }
                 \ra.
\label{C-}
\ee
The total irreducible Coulomb correction is the sum of
(\ref{CeN}), (\ref{Cee}), and (\ref{C-}):
\ba
\delta_{C} E &=& \fr{ E\pi\alpha }{ 4 m^3 }
         \la 4 \delta(\r)
         - 3 Z \left[ \delta(\r_1)+\delta(\r_2) \right] \ra
         \non \\
      && + \la \fr{ 3\pi \alpha Z }{ 4m^3 }
           \left[ \delta(\r_1)
            \lb \fr{ p_2^2  }{ 2m }
            + C_2 + c \rb
            + (1 \leftrightarrow 2) \right]
         - \fr{ \pi\alpha\delta(\r) }{ m^3 }
               \lb \fr{ 3P^2 }{ 32m } + C_{\rm N} \rb
               \right.
         \non \\
       &&
         \left.
         + 3  \fr{ \left[ \Pv \p,
        \left[ \Pv \p, c \right] \right] }{ 32m^4 }
         + \fr{ (\E_1-\E_2)\e }{ 32m^3 }
         - \fr{ {\cal E}_1^2+{\cal E}_2^2 }{ 32m^3 }
         + \fr{ e^2 }{ 8m^3 }
         \ra.
\label{coul}
\ea


\subsubsection{Magnetic corrections}

These corrections originate from single magnetic
exchanges between particles in instantaneous
approximation. There are two sources of such effects.
Relativistic corrections to the instantaneous
interaction of the Pauli currents of the electrons
induce the following contribution to the energy:
\ba
\delta_{M}^{\rm ee} E &=&
              \la \left\{
              \fr{ p_1^2+p_2^2 }{ 2m },
                 \fr{ \p_1 c\p_2 + (d-2)(\p_1\n)c(\n\p_2) }{4m^3}
               - \fr{ \pi\alpha\delta(\r) }{ m^3 }
                    \right\}
                \right. \non \\&& \left.~~~~
             - \fr{ d-1 }{ 32m^4 }
             \lb \left[ p_1^2,  \left[ p_1^2, c  \right] \right]
               + \left[ p_2^2,   \left[ p_2^2, c  \right] \right]
             \rb \ra,
\ea
which can be transformed to
\ba
\delta_{M}^{\rm ee} E &=& \fr{ E }{ m } \la \fr{ \p_1
c\p_2 + (\p_1\n)c(\n\p_2)}{2m^2} - \fr{ 2\pi\alpha
\delta(\r)
            }{ m^2 }  \ra
\non \\
&& + \la \fr{ 2\pi\alpha \delta(\r)
             }{ m^3 }  C_{\rm N}
    - \left\{ \fr{ C }{ 4m^2 },
       \fr{ \p_1 c\p_2 + (d-2)
       (\p_1\n)c(\n\p_2)}{m} \right\}
      \right.
\non \\
&&
\left. - \fr{ \left[ \Pv \p,
        \left[ \Pv \p, c \right] \right] }{ 8m^4 }
       + \fr{ (\E_1-\E_2)\e }{ 8m^3 }
       + \fr{ d-1 }{8}\fr{ e^2 }{ m^3 }  \ra.
\label{magn+}
\ea
Then, virtual transitions to negative-energy states of
one electron induced by a single magnetic exchange
with the other and a single Coulomb exchange with both
other particles, shift the energy by
\be
\delta_{M}^- E =  \la - \fr{ (\E_1-\E_2)\e  }{ 2m^3 }
                     - \fr{ d-1 }{ 2 }
                      \fr{ e^2 }{ m^3 } \ra.
\label{magn-}
\ee


\subsubsection{Retardation corrections}

The effect of retardation on a single magnetic
exchange between the electrons is described in the
appropriate order by the following average value:
\be
\delta_R E = \la \int \fr{d^d q}{(2\pi)^d}
                     \fr{ 2\pi\alpha }{ q^4 } \left[
                          H, e^{-i\q\r_1}\fr{\p_1^q
                          + [\vec{s}_1\q,\vec{s}_1]}{m}
                          \right]
                          \left[
                          H, e^{i\q\r_2}\fr{\p_2^q
                          - [\vec{s}_2\q,\vec{s}_2]}{m}
                          \right]
                      + {\rm H.c.} \ra,
\label{ret}
\ee
where $\q$ is the magnetic photon's momentum and
$\p^q_a = \p_a - \q(\p_a\q)/q^2$. The correction
(\ref{ret}) consists of zero-, single-, and
double-Coulomb parts:
\ba
\delta_R E &=& \Delta_0 + \Delta_1 + \Delta_2,
\\
\Delta_0 &=& \la \int \fr{d^d q}{(2\pi)^d}
                      \fr{ \pi\alpha }{ m^4 }
                           \left[ p_1^2 , \left[ p_2^2,
                           \fr{ \p_1^q \p_2 - \fr{d-1}{4} q^2 }{ q^4 }  e^{i\q\r} \right]\right] \ra,
\\
\Delta_1 &=& \la \int \fr{d^d q}{(2\pi)^d}
                 \fr{ 2\pi\alpha }{ m^3 }
                           \left[ C,\p_1 \right]
                           \left[ p_2^2, \fr{\p_2^q}{q^4}
                            e^{i\q\r} \right]
                           + (1\leftrightarrow 2) \ra,
\\
\Delta_2 &=& \la \int \fr{d^d q}{(2\pi)^d}
                 \fr{ 4\pi\alpha }{ m^2 }
                           \left[ C,\p_1^q \right]
                           \fr{e^{i\q\r}}{q^4}
                           \left[ C,\p_2 \right] \ra.
\ea
Integrating over $\q$ and using the Schr\"odinger
equation we transform these expressions to the
following ones:
\ba
\Delta_0 &=& - \fr{ E^2 \la c \ra }{ 8 m^2 }
             + \fr{ E \la Cc \ra }{ 4m^2 }
             + \la - \fr{ C_{\rm N}^2 c }{ 8m^2 }
             - \fr{ C_{\rm N} c^2 }{ 4 m^2 }
             + \fr{ \pi\alpha\delta(\r) }{ 2m^4 } P^2
             + \fr{\p c \p}{ 4m^4 } P^2
                       \right.
                      \non \\
               &&
             -  \fr{ (\Pv\p) c (\Pv\p) }{ 8m^4 }
             - \fr{p_1^2 c (\n\p_2)^2 + (\p_1\n)^2 c p_2^2
                           -
                           3(\p_1\n)^2 c (\n\p_2)^2
                           + (1\leftrightarrow 2)}{ 16 m^4 }
                      \non \\
               &&  \left.
                       + \fr{ (\E_1-\E_2)\e }{ 8 m^3 }
                      - \fr{ c^3 }{ 8m^2 }
                      + \fr{ d-1 }{ 8 } \fr{ e^2 }{m^3}\ra,
\label{ret0}
                      \\
\Delta_1 &=& \fr{E \la c^2 \ra}{4m^2}
               + \la - c r
                          \fr{ 2(\n\p_2)(\E_1\p_2)
                    + (\n\E_1)\left[(\n\p_2)^2
                    - p_2^2\right]}{16 m^3}
                    + {\rm H.c.}\right.
                        \non \\
               &&     \left.
                       - 3\fr{ (\p_1\n)^2 c^2 + c^2 (\n\p_1)^2 }{ 16 m^3 }
                       + (1\leftrightarrow 2)
                       - \fr{ C_{\rm N} c^2 }{ 4m^2 }
                       - \fr{d-2}{4} \fr{ c^3 }{ m^2 }
                       + \fr{d-1}{4} \fr{ e^2 }{ m^3 } \ra,
\label{ret1}
                        \\
\Delta_2 &=& \la c r^2
            \fr{ 3 \E_1\E_2 -
                (\n\E_1)(\n\E_2)
                - 2(\E_1-\E_2)\e }{8m^2}
                - \fr{(d-1)(d-2)^2}{8(4-d)} \fr{c^3}{m^2} \ra.
\label{ret2}
\ea
Here and below I use short-hand notations $(\n\p_a)^2$
and $(\p_a\n)^2$ for the operator $n_i(\n\p_a)p_{a i}$
and its hermitean conjugate, respectively.


\subsubsection{Seagull correction}

Double magnetic exchange between the two electrons one
of which goes over into negative-energy intermediate
states gives rise to the energy shift
\be
\delta_{\rm seag} E = \la \fr{ \p_1 c^2 \p_1
               + 3(\p_1\n) c^2 (\n\p_1)}{ 8m^3 }
               + (1\leftrightarrow 2)
               + \fr{ d-1 }{ 4 } \fr{ e^2 }{ m^3 }\ra.
\label{seag}
\ee
This correction completes the list of the ${\cal
O}(m\alpha^6)$ irreducible contributions to the helium
energy.


\subsection{Reducible Corrections}

\subsubsection{Breit Hamiltonian}

Breit Hamiltonian for the helium singlet states,
\be
U = U_S + U_P,
\ee
consists of two parts, $U_S$ and $U_P$, with the
selection rules $|\Delta {\cal S}|=0$ and $|\Delta
{\cal S}|=1$, respectively\footnote{$\vec{{\cal S}} =
\vec{s}_1 + \vec{s}_2$}. Just like in the positronium
case (see \cite{cmy99A}), the second order iteration
of the $S$-wave Breit perturbation diverges in three
dimensions. That is why this perturbation should be
considered in $d$ dimensions:
\be
U_S = - \fr{ p_1^4+p_2^4 }{ 8m^3 }
      + \fr{ \pi Z\alpha }{ 2m^2 }
        \lb \delta(\r_1) + \delta(\r_2) \rb
      + \fr{ (d-2) \pi\alpha }{ m^2 } \delta(\r)
 - \fr{ \p_1 c \p_2 + (d-2)(\p_1\vec{n})c(\vec{n}\p_2) }{ 2m^2
 }.
\label{USinit}
\ee
As for the $P$-wave part, which mixes singlet $S$ and
triplet $P$ states, the corresponding second order
iteration is saturated by the soft scale and therefore
we can take this perturbation in the limit of $d=3$:
\be
U_P =  \alpha \fr{ \vec{s}_1-\vec{s}_2 }{ 4m^2 }
        \lb \fr{ Z \vec{l}_1 }{ r_1^3 }
            - \fr{ Z \vec{l}_2 }{ r_2^3 }
            + \fr{ \r\times\Pv }{ r^3 } \rb.
\ee
Here $\vec{l}_a = \r_a \times \p_a$. Below I consider
energy shifts arising in second order in $U_P$ and
$U_S$.


\subsubsection{Second iteration of the Breit Hamiltonian:
$P$-wave}

To find the singlet $S$ level shift induced by an
admixture of triplet $P$ states,
\be
\delta_P E = \la U_P G U_P \ra,
\label{DPEinit}
\ee
where $G$ is the (reduced) Green function of the
Schr\"odinger equation, consider first  the action of
$U_P$ on the ground state wave function. As far as the
latter depends on the absolute values of $\r_1$,
$\r_2$ and $\r$, $\psi=\psi(r_1,r_2,r)$, we can
substitute:
\ba
\vec{l}_1 = \r_1\times\p_1 &\to&
              -i \r_1\times\lb \n_1 \partial_1 + \n\partial \rb
              = i   \fr{\r_1 \times \r_2}{r}
              \partial,
              \quad
\vec{l}_2 \to - i \fr{\r_1 \times \r_2}{r} \partial,
              \non \\
\r\times\Pv &\to& - i \r_1 \times \r_2
                  \lb \fr{1}{r_1}\partial_1
                  + \fr{1}{r_2}\partial_2 \rb,
              \non \\
U_P &\to& i\alpha \fr{ \vec{A}(\vec{s}_1-\vec{s}_2)}{
                     2m^2 }
              \lb  \fr{ Z }{ r_1^3 r } \partial
            - \fr{ 1 }{ r^3 r_1 } \partial_1
             + (1 \leftrightarrow 2)  \rb,
\ea
where $\partial_a = \partial/\partial r_a$, $\partial
= \partial/\partial r$, while the vector
\be
\vec{A} = \fr{ \vec{r}_1\times\vec{r}_2 }{ 2 }
         = \fr{ \vec{r} \times\vec{r}_1 }{ 2 }
         = \fr{ \vec{r} \times\vec{r}_2 }{ 2 }
\label{A}
\ee
is perpendicular to the triangle composed by $\r_1,
\r_2$ and $\r$, while its norm equals this triangle's
area. The $P$ wave admixture to the $S$ state wave
function,
\be
\delta_P \psi = G U_P \psi,
\ee
has the same angular dependence as the perturbation:
\be
\delta_P \psi =  i\alpha \fr{
                        \vec{A}(\vec{s}_1-\vec{s}_2)}{
                        2m^2 }
            \fr{ \phi_P(r_1,r_2,r) }{ r_1 r_2 r }.
\label{P-ansatz}
\ee
Function $\phi_P(r_1,r_2,r)$ is introduced in such a
way in order to make it finite at coalescense points:
$\phi_P(r_1,r_2,r)<\infty$ when $r_a \to 0$ or $r \to
0$. Substituting the r.h.s. of Eq.(\ref{P-ansatz})
into the inhomogeneous equation,
\be
(E - H)\delta_P \psi = U_P \psi,
\ee
we can cancel the spin- and angular-dependent factor
using the following relations:
\ba
\left[H,\vec{s}_1-\vec{s}_2 \right]
             &=& 0;\non \\
\left[H, \vec{r}_1\times\vec{r}_2 \right]
             &=&
             \left[ \fr{ p_1^2 }{ 2m }, \vec{r}_1
             \right] \times\vec{r}_2
             + \vec{r}_1\times
             \left[ \fr{ p_2^2 }{ 2m }, \vec{r}_2
             \right]
             \to- \fr{ \vec{r}_1\times\vec{r}_2 }{ m }
             \lb \fr{1}{r_1}\partial_1 + \fr{1}{r_2}\partial_2
             + \fr{2}{r}\partial\rb.    \non
\ea
At the last step I took into account that the operator
acts on the function depending only on $r_1,r_2$ and
$r$. The resulting equation for $\phi_P$ reads:
\be
\left\{H' - \fr{1}{ m }
 \lb \partial_1 \fr{1}{ r_1 } + \partial_2 \fr{1}{ r_2 } +
     \partial \fr{2}{ r } \rb
          - E  \right\} \phi_P
             = \left\{ \fr{ r_1 }{ r^2 } \partial_2
             - \fr{ Z r_1 }{ r_2^2 } \partial
             + (1 \leftrightarrow 2)\right\} \psi.
\ee
Here $H'$ is obtained from $H$ by the substitutions
$\partial \to
\partial - 1/r,
\; \partial_1 \to \partial_1 - 1/r_1, \; \partial_2
\to
\partial_2 - 1/r_2.$

The final expression for the energy shift
(\ref{DPEinit}) then reads:
\be
\delta_P E = \fr{ \alpha^2 }{ 4 m^4 }
             \la \phi_P \left|
             \fr{  A^2 }{  r^2 r_1^2 r_2^2 }
             \left\{ \fr{ r_1 }{ r^2 } \partial_2
             - \fr{ Z r_1 }{ r_2^2 } \partial
             + (1 \leftrightarrow 2)\right\}
            \right| \psi \ra.
\label{pbreit}
\ee


\subsubsection{Second iteration of the Breit
Hamiltonian: $S$-wave}

In order to extract divergences from the second order
correction induced by the $S$-wave part of the Breit
Hamiltonian, Eq.(\ref{USinit}), it is convenient to
rewrite the latter in the form
\be
U_S = - \fr{H^2}{2m} + \{ H, u \} - 2E u + {\cal O}_d,
\label{US}
\ee
where
\ba
u &=& \fr{ C_{\rm N} + (d-1)c }{4m},
\\
{\cal O}_d &=&   E \fr{ C_{\rm N} + (d-1)c }{2m}
                 + \fr{1}{8m}\left\{
                 \fr{p_1^2}{m} + c, \fr{p_2^2}{m} + c
                 \right\} - \fr{C_{\rm N}c}{2m}
                 - \fr{2d-3}{4m}c^2
                 \non
\\
&& + \fr{\p_1\lb C_{\rm N} - (d-2)c \rb\p_1
        + (1\leftrightarrow 2)}{4m^2}
        - \fr{ \p_1 c \p_2
        + (d-2)(\p_1\vec{n})c(\vec{n}\p_2) }{ 2m^2 }.
\ea
Inserting (\ref{US}) into equation
\be
\delta_S E = \la U_S G U_S \ra,
\label{DSEinit}
\ee
and using the Schr\"odinger equation for the reduced
Green function, $(H-E)G(R,R')=\psi(R)\psi(R') -
\delta(R-R')$, where $R$ denotes the vector
$(r_1,r_2,r)$, we get:
\be
\delta_S E = 2 \la U_S \ra \la u \ra
             - \la \fr{1}{2} \left[ [H,u],u \right]
             + \{ U_S, u \} \ra
             + \la {\cal O}_3 G {\cal O}_3 \ra.
\label{DSEexpan}
\ee
Note that we can take the limit $d \to 3$ in the last
term since in contrast to $U_S$ the perturbation
${\cal O}_3$ does not contain operators more singular
than $C_a^2$ and $c^2$. In other words, all the
divergences in (\ref{DSEexpan}) are moved to the
average values of local operators.

The first term from Eq.(\ref{DSEexpan}) is most easy
to calculate:
\be
2 \la U_S \ra \la {\cal O}_1 \ra = B
               \fr{2E + \la c \ra}{2m}.
\label{factorized}
\ee
Here $B = \la U_S \ra$ is the non-recoil limit of the
first order Breit correction (\ref{breitcorr}). For
the second term from Eq.(\ref{DSEexpan}) we have
\be
- \la \fr{1}{2} \left[
                  [ H,u],u \right] \ra
    = \la \fr{ {\cal E}_1^2+{\cal E}_2^2 }{ 32m^3 }
    + \fr{ (\E_1-\E_2)\e }{ 8 m^3 }
    + \fr{ (d-1)^2 e^2 }{ 16 m^3 } \ra.
\label{comm}
\ee
Then, the third term from Eq.(\ref{DSEexpan}) can be
rewritten as
\ba
- \la \{ U_S, u \} \ra &=&
 \fr{E^3}{2 m^2}
- \fr{E\la C^2 \ra}{ 2 m^2 } - \fr{\pi \alpha (Z-2
)}{4m^3}
  \la \delta(\r_1)C_2+\delta(\r_2)C_1 \ra
\non \\
&&
+ \la \fr{ 3 C c C_{\rm N} }{ 4 m^2 } - \fr{ p_1^2
C_{\rm N} p_2^2 }{ 8 m^4 } + \fr{\left[ \Pv \p,\left[
\Pv \p, c \right] \right]
   + 2 (\Pv\p) c (\Pv\p)}{8 m^4}
\right.
\non \\
&&
- \fr{ \pi\alpha\delta(\r)}{2m^3} C_{\rm N} + \left\{
\fr{C_{\rm N}+(d-1)c}{8m^2},
  \fr{\p_1 c\p_2 + (d-2)
  (\p_1\n)c(\n\p_2)}{m} \right\}
\non \\
&& \left. - \fr{ (\E_1-\E_2)\e }{
       4 m^3 }
       + \fr{C_{\rm N}^3}{4 m^2} + \fr{d-1}{8 m^2}
c^3 - \fr{ {\cal E}_1^2+{\cal E}_2^2 }{ 8 m^3 } - \fr{
(d-1) e^2 }{ 8 m^3 } \ra.
\label{antic}
\ea
Finally, the fourth term from Eq.(\ref{DSEexpan}) can
be calculated in a way similar to that used above for
the $P$-wave contribution. Namely, we first find the
solution for the inhomogeneous equation
$$
(E - H)\delta_S \psi = \lb {\cal O}_3
                       - \la {\cal O}_3 \ra \rb \psi,
$$
orthogonal to the ground state, $\la \delta_S \psi |
\psi \ra=0$, and then evaluate the matrix element of
${\cal O}_3$ between $\delta_S \psi$ and $\psi$:
\be
\la {\cal O}_3 G {\cal O}_3 \ra = \la \delta_S \psi
\left| {\cal O}_3 \right| \psi \ra.
\label{O3iter}
\ee
An alternative way to calculate (\ref{O3iter}) is to
include the operator ${\cal O}_3$ directly into the
Hamiltonian:
\be
(H + {\cal O}_3)\psi' = E'\psi'.
\ee
Then, with the ${\cal O}(\alpha^2)$ precision,
\be
\la {\cal O}_3 G {\cal O}_3 \ra =
                E' - E - \la {\cal O}_3 \ra.
\ee

\subsection{Total soft scale contribution}

Summing up all soft scale contributions, given by
Eqs.(\ref{disp}, \ref{coul}, \ref{magn+}, \ref{magn-},
\ref{ret0}--\ref{seag}, \ref{pbreit},
\ref{factorized}--\ref{O3iter}), we get:
\ba
\delta_{\rm soft} E &=& - \fr{E^3}{2m^2}
                        + \fr{E^2 \la c \ra}{4m^2}
            + \fr{E}{m}\la \fr{2C_{\rm N} C + c^2}{4m}
            - \fr{p_1^2 p_2^2}{8m^3}
            - \fr{\pi \alpha Z }{4m^2}
              \left[ \delta(\r_1)+\delta(\r_2)
              \right] \ra
             \non
             \\
          && + \fr{B \la c \ra}{2m}
             + \delta_P E
             + \la {\cal O}_3 G {\cal O}_3 \ra
             \non
             \\
          && + \la - \fr{c C_{\rm N} C}{ 2m^2 }
             - \left\{ \fr{ C_{\rm N} }{ 8 m^3 },
      \fr{\p_1 c \p_2 + (\p_1\n)c(\n\p_2)}{m} \right\}
             + \fr{p_1^2 C_{\rm N} p_2^2}{ 4m^4 }
             \right.
             \non
             \\
          &&
             + \fr{\p_1 c^2 \p_1 + \p_2 c^2 \p_2}{ 8m^3 }
             + \fr{\p c \p P^2 - (\Pv\p) c (\Pv\p)}{ 4m^4 }
             \non \\
         &&
             - \fr{p_1^2 c (\n\p_2)^2 + (\p_1\n)^2 c p_2^2
                           - 3 (\p_1\n)^2 c (\n\p_2)^2
                           + (1\leftrightarrow 2)}{ 16m^4 }
             \non \\
          && - \fr{c r}{16m^3}
               \lb 2(\n\p_2)(\E_1\p_2)
             + (\n\E_1)\left[(\n\p_2)^2-p_2^2\right]
             + (1\leftrightarrow 2) + {\rm H.c.} \rb
             \non
             \\
          && + \fr{c r^2}{ 8 m^2 }
               \lb 3 \E_1\E_2 - (\n\E_1)(\n\E_2)
                      - 2 (\E_1-\E_2)\e \rb
             \non
             \\
          && - \fr{ 3\left[ \Pv \p, \left[ \Pv \p,
                    c \right] \right] }{ 32 m^4 }
             + \fr{ \pi\alpha\delta(\r) }{ 2 m^3 }
               \lb \fr{ 13 P^2 }{ 16 m } + C_{\rm N} \rb
             \non
             \\
          && + \fr{\pi\alpha Z}{ 4m^3 }
               \left[ \delta(\r_1) \lb \fr{3p_2^2}{2m}
               + 2C_2 + c \rb
              + (1\leftrightarrow 2) \right]
              - \fr{ (\E_1-\E_2)\e }{ 32 m^3 }
             \non
             \\
          && \left.- \fr{ C_{\rm N}^3 }{ 4m^2 }
          + \fr{ {\cal E}_1^2+{\cal E}_2^2 }{ 8m^3 }
          - \fr{ 1 - 3\ep }{ 2 m^2 } c^3
          + \fr{ 3 - 6\ep }{ 4 m^3 } e^2 \ra.
\label{totsoft}
\ea
Order $\ep$ coefficients are kept in
Eq.(\ref{totsoft}) only if they multiply the operators
whose average values contain $1/\ep$ singularities. In
(\ref{totsoft}), the following relations are taken
into account:
\be
[p_i,[p_j,[c^2 n_i n_j]] = \fr{3-d}{d-2} e^2,~~~~~
(d-3) \la \left\{ c,
      \fr{\p_1 c \p_2 + (\p_1\n)c(\n\p_2)}{m} \right\} \ra
    = (d-3) \la c^3 \ra + {\cal O}(\ep).
\ee
For the bulk of the operators in (\ref{totsoft}),
their average values can be safely evaluated at $d \to
3$. Special care is needed when one deals with the
operator $p_1^2 c (\n\p_2)^2 + (1\leftrightarrow 2) +
{\rm H.c.}$ which is not well defined in three
dimensions. We can take the limit $d \to 3$ having
previously determined what the momenta operators are
acting on:
\be
\lim_{d \to 3} \la p_1^2 c (\n\p_2)^2 +
              (1\leftrightarrow 2) + {\rm H.c.} \ra =
\left. \la \overleftarrow{p_1^2} c
           \overrightarrow{(\n\p_2)^2} +
              (1\leftrightarrow 2) + {\rm H.c.}
              + 4\pi m^2 \alpha^3 \delta(\r) \ra
              \right|_{d=3}.
\label{anom}
\ee
In order to be certain that this relation holds, it is
sufficient to consider the two-body problem, where
both sides of Eq.(\ref{anom}) can be calculated
analytically.

In order to calculate singular average values entering
into (\ref{totsoft}), consider the Coulomb potential
$C_{ab}$ between two particles with charges $z_a$ and
$z_b$. The average value of $C_{ab}^3$ is
\be
\la C_{ab}^3 \ra = (z_a z_b \alpha)^3
    \lb \fr{\Gamma \lb \fr{d}{2}-1 \rb }{ \pi^{d/2-1} } \rb^3
    \la r_{ab}^{6-3d} \ra.
\ee
Repeating the procedure used in Sec. \ref{ma5} we get
\be
\la C_{ab}^3 \ra = (z_a z_b\alpha)^3 \left\{
    \lb \fr{1}{\ep} - 4\ln \fr{2}{{\rm a}_{ab}} + 2 \rb
    \la \pi\delta(\r_{ab}) \ra
  - 2 \la \fr{ \gamma + \ln\fr{2 r_{ab}}{{\rm a}_{ab}}
             }{ r_{ab}^2 }
     \n_{ab} \overrightarrow{\gra}_{ab} \ra
      \right\}.
\label{c3}
\ee
Recall that ${\rm a}_{ab} = |z_a z_b
\mu_{ab}\alpha|^{-1} $. Average value of the second
singular operator, ${\cal E}_{ab}^2$, can be found in
the following way:
\be
\la {\cal E}_{ab}^2 \ra = \la \left[ \gra_a, C_{ab}
                     \right]^2 \ra
                   = \la \left[ \gra_a, C_{ab}
                     [ \gra_a, C_{ab} ] \right]
                   - C_{ab} [ \gra_a,
                     [ \gra_a, C_{ab} ]] \ra
                   = -2 \la C_{ab}[ \gra_a,
                        C_{ab}] \overrightarrow{\gra}_a \ra.
\ee
At the last step, I again have used the equation $\la
C_{ab} \delta(\r_{ab})\ra=0$ valid in the dimensional
regularization. In order to express the average value
\be
\la {\cal E}_{ab}^2 \ra = 2(d-2)(z_a z_b \alpha)^2
 \lb \fr{\Gamma \lb \fr{d}{2}-1 \rb }{ \pi^{d/2-1} } \rb^2
    \la r^{3-2d} \n_{ab} \overrightarrow{\gra}_{ab} \ra
\ee
in terms of $\la C_{ab}^3 \ra$, the known derivative
of the wave function at $\r_{ab} \to
0$\footnote{Strictly speaking, with a coefficient $1 +
{\cal O}(\ep)$. Inspection of the known two-body
average values, however, shows that the coefficient is
in fact equal to 1.} can be added to and subtracted
from $\n_{ab} \overrightarrow{\gra}_{ab}$:
\be
\la {\cal E}_{ab}^2 \ra = 2\mu_{ab}(d-2)\la
            C_{ab}^3 \ra
          + 2(z_a z_b \alpha)^2
       \la \fr{1}{r_{ab}^3}
       \lb \n_{ab} \overrightarrow{\gra}_{ab}
        - \mu_{ab} z_a z_b \alpha \rb \ra.
\label{e2}
\ee
Since the last average value in the r.h.s. of
(\ref{e2}) is finite, it can be considered in three
dimensions. Now, extracting all divergent pieces from
(\ref{totsoft}) we get (\ref{softdiv}).

\section{Conclusion}
\label{summ}

The divergences contained in the hard (\ref{rec}) and
soft (\ref{softdiv}) scale contributions cancel each
other so that in the sum of  all contributions,
Eqs.(\ref{radrec}--\ref{rec}, \ref{totsoft}), we can
put $d = 3$. Taking into account
Eqs.(\ref{anom},\ref{c3},\ref{e2}), the final
expression for the ${\cal O}(m\alpha^6)$ correction to
a singlet $S$-state energy of the helium atom can be
written as
\ba
\delta^{(4)} E &=& - \fr{E^3}{2m^2}
                        + \fr{E^2 \la c \ra}{4m^2}
            + \fr{E}{m}\la \fr{2C_{\rm N} C + c^2}{4m}
            - \fr{p_1^2 p_2^2}{8m^3}
            - \fr{\pi \alpha Z }{4m^2}
              \left[ \delta(\r_1)+\delta(\r_2)
              \right] \ra
             \non
             \\
          && + \fr{B \la c \ra}{2m}
             + \delta_P E
             + \la {\cal O}_3 G {\cal O}_3 \ra
             + \pi \alpha^3 m^2 \la k_{\rm eN}
              \lb \delta(\r_1)+\delta(\r_2) \rb
             + k_{\rm ee} \delta(\r) \ra
             \non
             \\
          && + \la
             - \fr{3 C_1 C_2 C_{\rm N}}{ 4m^2 }
             - \fr{c C_{\rm N} C}{ 2m^2 }
             - \fr{ C_{\rm N} c [\p_1 \p_2
             + \n(\n\p_1)\p_2] + {\rm H.c.}}{8m^4}
             \right.
             \non
             \\
          &&
             + \fr{p_1^2 C_{\rm N} p_2^2}{ 4m^4 }
             + \fr{\p_1 c^2 \p_1 + \p_2 c^2 \p_2}{ 8m^3 }
             + \fr{(\p_1\times\p_2) c (\p_1\times\p_2)}{ 4m^4 }
             \non \\
         &&
             - \fr{p_1^2 c (\n\p_2)^2 + (\p_1\n)^2 c p_2^2
                           - 3 (\p_1\n)^2 c (\n\p_2)^2
                           + (1\leftrightarrow 2)}{ 16m^4 }
             \non \\
          && - \alpha\fr{ 2(\n\p_2)(\E_1\p_2)
             + (\n\E_1)\left[(\n\p_2)^2-p_2^2\right]
             + (1\leftrightarrow 2) + {\rm H.c.} }{16m^3}
             \non
             \\
          && + \alpha r\fr{3 \E_1\E_2 - (\n\E_1)(\n\E_2)
                      - 2 (\E_1-\E_2)\e}{ 8 m^2 }
             - \fr{ 3\alpha }{ 32 m^4 }
              \fr{ P^2-3(\n\Pv)^2 }{ r^3 }
               \non
             \\
          && + \fr{ \pi\alpha\delta(\r) }{ 2 m^3 }
               \lb \fr{ 9 P^2 }{ 16 m } + C_{\rm N} \rb
             + \fr{\pi\alpha Z}{ 4m^3 }
               \left[ \delta(\r_1) \lb \fr{3p_2^2}{2m}
              - \fr{(2Z-1)\alpha}{r_2} \rb
              + (1\leftrightarrow 2) \right]
              \non
             \\
          && - \fr{ (\E_1-\E_2)\e }{ 32 m^3 }
             + \fr{(Z \alpha)^2 }{ 4m^3 }
               \left[ \fr{1}{r_1^3} \lb \n_1\gra_1
             + m Z \alpha \rb
             + (1\leftrightarrow 2) \right]
            \non
             \\
          && \left.- \alpha^3 \fr{\ln(m\alpha r)
                   + \gamma}{2m^2 r^2}\n \gra
         + \fr{3\alpha^2 }{ 2m^3 }
           \fr{1}{r^3} \lb \n \gra - \fr{m\alpha}{2} \rb \ra.
\label{tot6}
\ea
Here all momentum operators standing to the right
(left) of position-dependent operators are assumed to
act on the right (left) wave function. Although
$d$-dimensional notations for the Coulomb potentials
and electric forces are kept in (\ref{tot6}), the
immediate three-dimensional counterparts are implied
for all operators, e.g., $C_1 \to -Z\alpha/r_1$, $\e
\to \alpha\n/r^2$ and so on. The contact terms enter
into Eq.(\ref{tot6}) with the coefficients
\ba
k_{\rm eN} &=& \fr{Z^3}{2} + \fr{427 Z^2}{96} - \fr{10
Z}{27}
         - \fr{9 Z \zeta(3)}{4\pi^2}
         - \fr{2179 Z}{648\pi^2}
         + \fr{3Z - 4Z^2}{2}\ln 2,\\
\label{ken}
k_{\rm ee} &=& - \ln\alpha + \fr{3385}{216} -
\fr{331}{54\pi^2}
         - \fr{29\ln 2}{2} + \fr{15 \zeta(3)}{4\pi^2}.
\label{kee}
\ea
Eqs.(\ref{tot6}-\ref{kee}) are the principal result of
the present work. Its application to the ground state
of the helium atom is considered elsewhere \cite{ky}.

The approach elaborated here can be applied to other
few-electron atoms as well as to higher order
corrections. The only stumbling block to higher order
calculations is the yet unknown three-loop hard scale
electron-electron potential. Nevertheless, the order
$m\alpha^7$ corrections enhanced by powers of $\ln
\alpha$ can be determined by combination of methods
used in Ref. \cite{my-logs} and in the present work.

\subsection*{Acknowledgments}

Useful advices from V.~L. Chernyak, A. Czarnecki, V.
Korobov, and especially K. Melnikov are gratefully
acknowledged. This research was supported in part by
the Russian Ministry of Higher Education and by the
Russian Foundation for Basic Research under grant
number 00-02-17646.

\end{document}